\documentclass[12pt]{article}
\usepackage{graphicx}
\usepackage{dcolumn}
\usepackage{bm}
\usepackage{amsfonts}
\usepackage{amsthm}
\usepackage{amsmath}

\usepackage{amssymb}
\usepackage{epsfig}
\usepackage{color}
\usepackage{textcomp}

\usepackage{hyperref}
\usepackage[retainorgcmds]{IEEEtrantools}
\usepackage{cite}

\def\be{\begin{equation}}
\def\ee{\end{equation}}
\def\ba{\begin{eqnarray}}
\def\ea{\end{eqnarray}}

\def\bdm{\begin{displaymath}}
\def\edm{\end{displaymath}}

\def\bq{\begin{quote}}
\def\eq{\end{quote}}

 at 10truept

\newcommand{\bea}{\begin{eqnarray}}
\newcommand{\eea}{\end{eqnarray}}

\newcommand{\bi}{\begin{itemize}}
\newcommand{\ei}{\end{itemize}}

\newcommand{\beq}{\begin{equation}}
\newcommand{\eeq}{\end{equation}}
\newcommand{\beqa}{\begin{eqnarray}}
\newcommand{\eeqa}{\end{eqnarray}}

\def\ltap{\ \raise.3ex\hbox{$<$\kern-.75em\lower1ex\hbox{$\sim$}}\ }
\def\gtap{\ \raise.3ex\hbox{$>$\kern-.75em\lower1ex\hbox{$\sim$}}\ }
\def\gl{\ \raise.5ex\hbox{$>$}\kern-.8em\lower.5ex\hbox{$<$}\ }
\def\roughly#1{\raise.3ex\hbox{$#1$\kern-.75em\lower1ex\hbox{$\sim$}}}

\begin{document}

\thispagestyle{empty}
\begin{flushright}
February 2018
\end{flushright}
\vspace*{1.2cm}
\begin{center}
{\Large \bf Quantum Cosmic No-Hair Theorem and Inflation}\\

\vspace*{1.7cm} {\large Nemanja Kaloper\footnote{\tt
kaloper@physics.ucdavis.edu} and James Scargill\footnote{\tt
jhcscargill@gmail.com}}\\

\vspace{.3cm} {\em Department of Physics, University of
California, Davis, CA 95616, USA}\\

\vspace{1.5cm} ABSTRACT
\end{center}

We consider implications of the quantum extension of the inflationary no hair theorem. We show that when the quantum state of inflation is picked to ensure the
validity of the EFT of fluctuations, it takes only ${\cal O}(10)$ efolds of inflation to erase the effects of the initial distortions on the inflationary observables. Thus the Bunch-Davies vacuum is a very strong quantum attractor during inflation. We also consider bouncing universes, where the initial conditions seem to linger much longer and the 
quantum `balding' by evolution appears to be less efficient.

\vfill \setcounter{page}{0} \setcounter{footnote}{0}
\newpage

\section{Introduction}

Inflation is a simple and controllable framework for describing the origin of the universe. It relies on rapid cosmic expansion and subsequent small
fluctuations described by effective field theory (EFT). Once the slow roll regime is established, rapid expansion wipes out random and largely undesirable initial features of the universe, and the resulting EFT of fluctuations on the expanding background replaces them with small, nearly scale invariant spectra of
scalar and tensor fluctuations. 
The beginning of inflation might be described by some of the existing theories, such as no-boundary, tunneling from nothing, pre-inflationary origin, or eternal inflation in the multiverse.\footnote{These ideas have been discussed so much in the literature that even identifying the proper original references at this point is difficult, to say the least. We shall only mention one of the earliest suggestions about quantum, uncertain, cosmic origin \cite{Tryon:1973xi}.} The exact details are largely irrelevant for the last 60 efolds when the observable features are generated, except for the description of the initial state. ``Common wisdom" dictates that the initial state is taken to be the Bunch-Davies vacuum,
which readily yields a spectrum of scale invariant perturbations. Yet picking this state ``by choice" can lead to confusion, and a critic might even object that this is a tuning, i.e., putting in the answer by handpicking the initial state. Further, there are concerns that the
fluctuation modes seem to appear out of nowhere in the Bunch-Davies vacuum, with apparently negligible initial
backreaction. 

As we will explain this is resolved by a proper application of EFT to fluctuations. First off, the real vacuum of the theory is the Bunch-Davies state \cite{Bunch:1978yq}.  This follows from the quantum no-hair theorem for de Sitter space, which selects the Bunch-Davies state as the vacuum with the UV properties that ensure cluster decomposition \cite{Marolf:2010zp, Marolf:2010nz,Hollands:2010pr}. The excited states on top of it obey the constraints arising from backreaction, to ensure that EFT holds \cite{Schalm:2004qk, Greene:2004np, Porrati:2004gz, Porrati:2004dm, Nitti:2005ym, Flauger:2013hra, Ashoorioon:2013eia, Jiang:2015hfa, Jiang:2016nok}. 
The initial state need not be the vacuum, but rather some deformation of it with some (quantum, as well as classical) memory of the initial conditions, whatever those may be.

Starting with this, 
we will quantify explicitly how quickly inflationary evolution wipes out the initial excitations and evolves the initial state to the point where it is {\it practically indistinguishable} from the Bunch-Davies vacuum. In other words, we will calculate the rate of the `thermalization' process induced by inflation on the quantum state of the universe.
This will generically require an additional ${\cal O}(10)$ efolds, during which the initial
excitations, random or entangled, will be reduced to be subleading to the intrinsic inflationary fluctuations in the
Bunch-Davies vacuum that generate the CMB anisotropies and serve as the seeds for structure formation. Moreover, we will also see that the same thermalization dynamics of the quantum vacuum reduces the initial nonlinearities, implying that the initial non-Gaussianities also diminish in the course of inflation, and that the non-Gaussianities which survive the initial $\sim 10$ efolds are really due to the nonlinear effects in the inflaton sector rather than initial conditions. Our results will explicitly show how the Bunch-Davies state is dynamically realized by inflation, and determine the point after which the standard calculations apply. Hence there is a price to pay for using the Bunch-Davies state for the computation of $\delta \rho/\rho$. The good news is that it is acceptable. Further, since the standard EFT is valid throughout this regime, there is no transplanckian problem whatsoever: it is merely a mirage that follows from inconsistent assumptions.

A similar argument used for the vacuum state employed in bouncing cosmologies generically shows that imposing the Bunch-Davies vacuum there is much costlier, since one must impose that the quantum
state is initially homogeneous over many more orders of magnitude. Bouncing cosmology models therefore need to include a mechanism which explains how the quantum vacuum was achieved.

\section{Transplanckian Addling and Cisplanckian Sangfroid}

As a starting point we briefly review the ``standard" description of the genesis of inflationary perturbations with a particular focus on the so-called ``transplanckian problem" \cite{Martin:2000xs}. We then point out a simple, logical resolution of the alleged problem, consistent with the framework of the EFT of quantum fluctuations of the inflaton and the cosmic no-hair theorem. 

Indeed imagine an inflationary theory in the slow roll regime, and truncate the theory of the fluctuations to only the Gaussian sector. Focusing for
simplicity only on the scalar field, in the longitudinal gauge one finds
\be
ds^2 = a^2\Bigl(-(1-2\Phi) {\rm d}\eta^2 + (1+2\Phi) {\rm d}\vec x^2\Bigr) \, , ~~~ \phi = \phi_0(\eta) + 
\delta \phi(\eta,\vec x) \, ,   ~~~
\dot \phi_0 \delta \phi = - 2 M_{\rm Pl}^2 \Bigl(\dot \Phi + \frac{\dot a}{a} \Phi \Bigr) \, ,
\label{background}
\ee
where $a$ is the scale factor in a flat FRW universe, and $\Phi$ is its scalar, Newtonian, perturbation generated by the
inflaton perturbation $\delta \phi$, because of the mixing induced by the time derivative of the inflaton background
$\dot \phi_0(\eta)$. Clearly, the system has only one degree of freedom, since once $\delta \phi$ is given, $\Phi$ is completely fixed. Turning to the dynamics of this degree of freedom, it is extremely convenient to use Mukhanov's curvature perturbation variable 
$\varphi = a \delta \phi - \frac{a \dot \phi_0}{\dot a/a} \Phi$, which obeys a free field equation in the FRW background \cite{Mukhanov:1981xt}. In momentum space,
\be
\ddot \varphi_k + \Bigl( \vec k^2 - \frac{2+\cdots}{\eta^2} \Bigr) \varphi_k = 0 \, ,
\label{free}
\ee
where the ellipsis denotes the slow roll corrections. Since roughly $\Phi \sim \varphi/a$, we can follow the time evolution of the fluctuations by using the solutions of the free field equation (\ref{free}). Using the physical wavelength of the modes, $\lambda(\eta) = \lambda_0 a(\eta) = a(\eta)/k$,  it is easy to see that 
\be
\Phi_k \simeq
\begin{cases}
\frac{\hat A}{a} \cos(k\eta + \delta) \, , ~~~ {\rm for} ~~ \lambda < 1/H \, ; \\
A + \frac{B}{a^3} \, , ~~~~~~~~~~~~ {\rm for}~~  \lambda > 1/H \, ,
\end{cases} \label{amplitude}
\ee
where $H$ is the (nearly constant) Hubble parameter during inflation. The latter case describes the inflationary 
freeze-out of perturbations. The former case describes the evolution of fluctuations at subhorizon scales, {\it ignoring} their interactions. The normalization coefficient $A$ can be calculated using EFT and declaring 
$A$ to be the expectation value of the inflaton's propagator in the Bunch-Davies state.

\begin{figure}[tp]
	\centering
	\includegraphics[width=13cm]{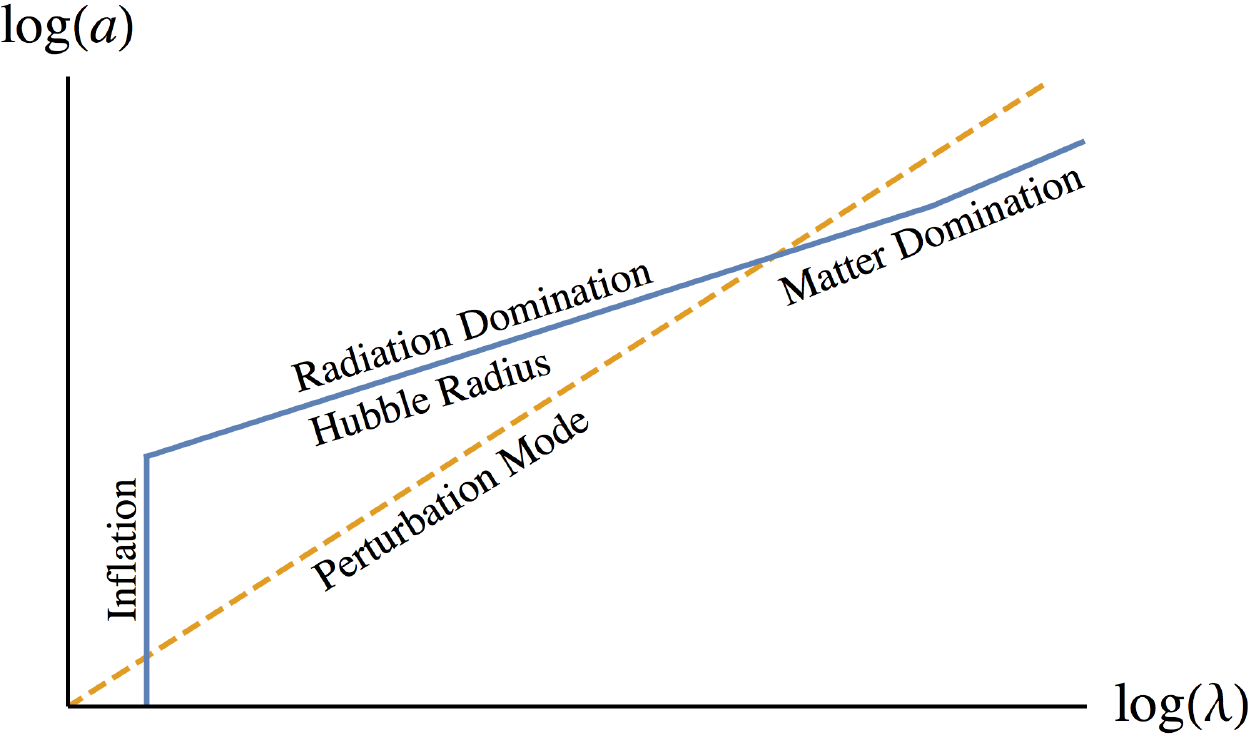}
	\caption{Wavelength of a fluctuation as a function of time during slow roll inflation.}
	\label{wavelength}
\end{figure}

Now, the statement of the ``transplanckian problem" \cite{Martin:2000xs} is that if one traces the fluctuations
back in time, from the horizon crossing to the earlier stages of inflation, one finds that the wavelength
shrinks exponentially, and before one blinks it will become shorter than the Planck length. Indeed, one can plot this
as in Fig. \ref{wavelength}.  This is then taken to mean that in order to really retain the predictivity of inflation as a means for determining the late-time amplitude of  fluctuations, one must specify the theory all the way to arbitrarily short lengths. Further, this might make one suspect that the inflationary results---specifically the scale invariance of $A$---might be a consequence of `fine tunings' hidden in the choice of the Bunch-Davies vacuum for the
EFT of fluctuations, which ``evidently" must be sensitive to ``transplanckian" physics.

\begin{figure}[tp]
	\centering
	\includegraphics[width=13cm]{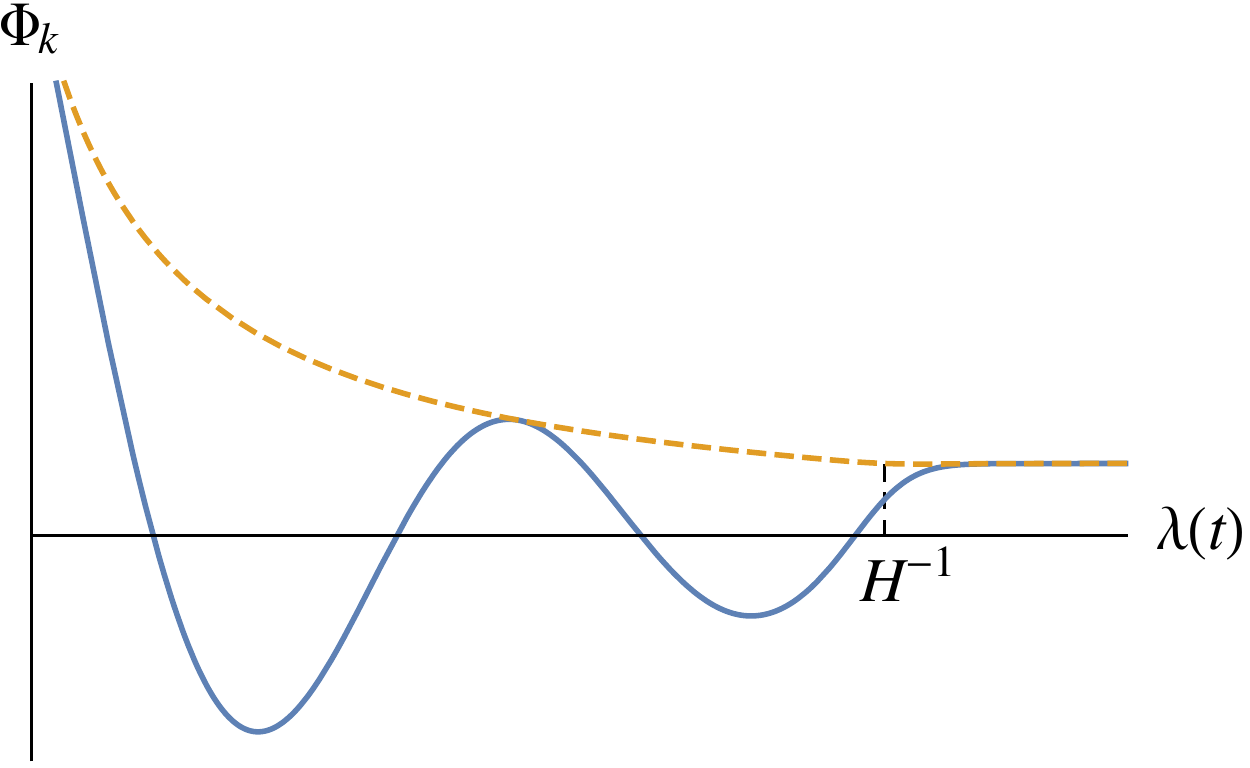}
	\caption{Amplitude of a fluctuation as a function of its wavelength (and thus time) during slow roll inflation.}
	\label{amplitudefig}
\end{figure}

This conclusion is faulty because it ignores the evolution of the amplitude of fluctuations for a given {\it physical} wavelength. 
Before horizon crossing, 
$\Phi_k \sim 1/a \sim 1/\lambda$ (as also plotted in Fig. \ref{amplitudefig}). This is just the virial theorem, since for $\lambda < 1/H$ the degrees of freedom of the scalar behave just like free harmonic oscillators in a cavity. Hence, if one extrapolates $\lambda$ back in time, one finds that its amplitude grows large! If one takes the scale of inflation as the highest allowed that might fit the data, by taking into account the bounds on primordial tensor modes,
$H < 10^{13} \, {\rm GeV}$, and fixing the amplitude $\Phi$ to $10^{-5}$ when $\lambda \simeq 1/H$ one finds that 
$\Phi \sim 1$ for $\lambda > 10 \, \ell_{\rm Pl}$. 

This means that linear perturbation theory in the Bunch-Davies vacuum cannot be extrapolated to transplanckian scales.
When $\lambda$ is short, gravitational effects become 
strong and the fluctuations distort the background dramatically. The linear approximation fails, and one cannot pass through this regime at one's whim \cite{Tanaka:2000jw}. The large blueshift factors enhance EFT interactions and the irrelevant operators cannot be ignored any more. 
In fact, even the background metric (\ref{background}) does not really make sense for these fluctuations: $\Phi$ is really an expectation value of the metric perturbation in the quantum state of inflation, and its dispersion will be too large. The dynamics at short distances yields all kinds of background distortions saturating with formation and evaporation of small black holes, which play the role of the dynamical cutoff. Hence, at the shortest scales some regions will be perturbed so much that they collapse and inflation never even starts there. In other regions of space, however, high energy dynamics would be less destructive and inflation can
begin \cite{East:2015ggf}. These regions would initially not be as smooth and flat, but inflation would iron them out as it goes, at the classical and quantum level. So a part of the transplanckian confusion is the misidentification of the quantum expectation value of the operator $\hat \Phi$ in an arbitrary state with the classical mode function $\Phi$ in the Bunch-Davies state. While this is correct in the Bunch-Davies state, it is not true in a general state where fluctuations are large. 

Instead,  the curvature fluctuations come about from particle production in an external field \cite{blokhintsev}, analogously to particle production in background fields in quantum electrodynamics, where pairs can be created by a large electric field in a parallel plate capacitor.
Pair creation will discharge the plates and decrease the field. 
Inflationary perturbations are similar, starting as the quanta of the inflaton placed on shell by the background fields
and perturbing the metric by generating 
density perturbations a long time after inflation, when the curvature perturbations of the geometry yield the perturbations of the matter density generated at reheating. 
The production rate 
will be slow thanks to a flat potential with a value significantly below the Planck scale, and slightly inhomogeneous because of the fluctuations. 
As the fluctuations evolve in the background, they will decohere, and 
soon after the fluctuation length becomes $\sim 1/H$ they
will cross the horizon and freeze. Clearly the fluctuations are never really subject to 
any significant ``transplanckian" influences. Simply put they never really probe physics beyond the Planck cutoff in a significant manner, thanks to decoupling. 

This picture will lay the foundation for our analysis that follows. Let us outline it: we will start with some generic initial
quantum state that is a non-trivial excitation of the Bunch-Davies state, which is selected by the cosmic no-hair theorems as the vacuum state of the EFT. We will then quantify the magnitude of the excitation by computing the backreaction in this state,  using the method of boundary actions as a clear-cut means to locally parametrize the initial deviations from Bunch-Davies. These are bounded so that the weakly coupled EFT is valid. This, by default, includes all those initial states where inflation can begin. Prior to this stage, the dynamics in the excited state was in strong coupling, and so it is not directly calculable without details of the full theory. Some states 
might not survive for long enough to allow inflation to set in; however, some will. The precise statement of the survival probability of a completely generic state is beyond the scope of this work. We will simply analyze the dynamics in the states that did survive, as those indeed support the weakly coupled EFT and are covered by our analysis. We will then study how long the initial deviations from  Bunch-Davies survive, and show that in generic states they become negligible after ${\cal O}(10)$ efolds. We will also find the magnitude of their corrections to the inflationary perturbations. Finally, we will consider similar processes in bouncing cosmologies.

\section{Boundary Action and Initial States}

Here we first review the formalism for implementing the quantum boundary conditions on the state of inflation which ensures the validity of the 
EFT description of fluctuations of the inflaton. This provides a systematic tool to parametrize the deviations of the initial state from the Bunch-Davies vacuum. Following \cite{Schalm:2004qk}, the basic idea is to supplement the action for the field whose perturbation we are considering with a term evaluated on a space-like boundary which encodes the initial conditions for the field. In particular, for a massless scalar\footnote{A good leading order approximation for the fluctuations of a light inflaton in slow roll.} one has
\begin{equation}
S = \frac{1}{2}\int \mathrm{d}^4x \sqrt{-g}\, \partial_\mu \phi \partial^\mu \phi + \frac{1}{2}\int_\Sigma \mathrm{d}^3x\mathrm{d}^3y \sqrt{\gamma(x)}\sqrt{\gamma(y)}\, \phi(x) \kappa(x,y) \phi(y) \, ,
\end{equation}
where $\gamma$ is the induced metric on the space-like surface $\Sigma$, and $\kappa(x,y)$ encodes information about the initial state of $\phi$. For a translation invariant state one has $\kappa = \kappa(|x-y|)$. Variation of the action with respect to $\phi$ (demanding, of course, that the variation is \emph{not} set to zero on the boundary, 
because the state initially differs from Bunch-Davies) yields the usual bulk equation of motion, along with the boundary condition
\begin{equation}
(\partial_n + \kappa) \phi = 0 \, , \label{boundary condition}
\end{equation}
where $\partial_n \equiv n^\mu \partial_\mu$ is the derivative normal to the boundary. Fourier transforming, and expanding the field in terms of creation and annihilation operators,
\begin{equation}
\phi(x,\eta) = \int \frac{\mathrm{d}^3k}{(2\pi)^3} \left( \varphi_-(\eta) A(k) + \varphi_+(\eta) A^\dagger(-k) \right) e^{i \mathbf{k} \cdot \mathbf{x}}\, ,
\end{equation}
the mode function $\varphi_+$ obeys \eqref{boundary condition}.
Reference \cite{Schalm:2004qk} emphasized that the physics is encoded in the mode functions, and so the precise location of the boundary is arbitrary, so long as one also changes $\kappa$ so as to ensure that \eqref{boundary condition} is still satisfied. Thus, we can fix the boundary at the conformal time $\eta_0$, and so $\partial_n = \partial_\eta/a$. Let us now use $\varphi_\pm$ to denote the mode functions in the Bunch-Davies vacuum, and denote their counterparts in more general states by 
$\varphi_b$.

The massless scalar field modes are the Hankel functions, $\varphi_\pm = \frac{H}{\sqrt{2k^{3/2}}} (1 \pm i k \eta ) e^{\mp i k \eta}$, 
and so the effective `interaction' of the modes in the Bunch-Davies state is given by
\begin{equation}
\kappa_{\text{BD}} = \frac{-k^2 \eta_0}{1 - i k \eta_0}\, .
\end{equation}
We will treat the corresponding quantities in a general state as a perturbation of the Bunch-Davies operators. This yields
\begin{equation}
\kappa = \kappa_\text{BD} + \delta \kappa \, , \qquad \varphi_b = \frac{1}{1-|b|^2} \left( \varphi_+ + b \varphi_- \right) \, . \label{modified state}
\end{equation}
Using \eqref{boundary condition} and the Klein-Gordon normalization of the mode functions $\varphi_+ \partial_n \varphi_- - \varphi_- \partial_n \varphi_+ = \frac{i}{a^3}$, the `Bogoliubov rotation' $b$ is given by
\begin{equation}
b = - \frac{\kappa \varphi_+ + \partial_n \varphi_+}{\kappa \varphi_- + \partial_n \varphi_-} \bigg|_{\eta=\eta_0} = \frac{i a_0^3 \delta\kappa \varphi_+^2}{1 - i a_0^3 \delta\kappa |\varphi_+|^2} \bigg|_{\eta=\eta_0} \, . \label{b coefficient}
\end{equation}
Then, a straightforward calculation shows that with the modes defined by \eqref{modified state} the Green function is
\begin{IEEEeqnarray}{rCl}
G_b(k; \eta_1, \eta_2) &=& \langle b | \phi_b(k, \eta_1) \phi_b(k, \eta_2) | b \rangle = \varphi_b^*(\eta_1) \varphi_b(\eta_2) ~~~~~ \nonumber \\
&=& G_b^{(0)} - \frac{H^4 a_0^3}{2 k^6} \mathrm{Im} \left[ \delta\kappa (1+ik\eta_0)^2 (1-ik\eta_1)(1-ik\eta_2) e^{-ik(2\eta_0 - \eta_1 - \eta_2)} \right]  \nonumber \\
&& \negmedspace {} + \frac{H^6 a_0^6}{8 k^9} \bigg[ |\delta\kappa|^2 (1+k^2\eta_0^2)^2 \left[ (1-ik\eta_1)(1+ik\eta_2) e^{-ik(\eta_1 - \eta_2)} + 2(\eta_1 \leftrightarrow \eta_2) \right]  \label{greenfunction} \\ 
&& \negmedspace {} - 2 \mathrm{Re}\left( \delta\kappa^2 (1+k^2\eta_0^2)(1-ik\eta_0)^2(1-ik\eta_1)(1-ik\eta_2) e^{-ik(2\eta_0 - \eta_1 - \eta_2)} \right)\bigg] + \mathcal{O}(\delta\kappa^3) \, , \nonumber
\end{IEEEeqnarray}
where $G_b^{(0)}$ is the Green function in the Bunch-Davies state.

\subsection{Backreaction}

The above mathematical framework can be used to estimate the backreaction of the excitations in the initial state on the evolution. Specifically,
the backreaction must be subleading to the background effects in controlling the evolution. 
A simple way to proceed is to consider the expectation value of the stress-energy tensor in the excited state as a leading order measure of backreaction; this follows from Ehrenfest's theorem. The nonzero value of 
$\delta \langle b | T^0_0 | b \rangle$ in the excited state must not overwhelm the background sources and impede inflation. The boundary action framework above then allows us to compute this quantity unambiguously and in 
a generic excitation of Bunch-Davies.

For a scalar 
field, the stress-energy tensor is $T_{\mu\nu} = \partial_\mu \phi \partial_\nu \phi - \frac{1}{2} g_{\mu\nu} \partial_\lambda \phi \partial^\lambda \phi$, and so\footnote{Note that since the Feynman two point function is the time ordered version of 
\eqref{greenfunction}, there is an extra factor of two in the integral for $\delta T$.}
\begin{equation}
\delta \langle b | T^0_0 | b \rangle = -\delta \langle b | T^i_i | b \rangle = -\frac{1}{a^2} \int \frac{\mathrm{d}^3k}{(2\pi)^3} \left( \partial_{\eta_1} \partial_{\eta_2} + k^2 \right) \delta G_b(k; \eta_1, \eta_2) \Big|_{\eta_1 = \eta_2 = \eta} \, . \label{delta T}
\end{equation}
A natural IR cutoff regulating the integral is given by $a H$, since comoving momenta smaller than this have already exited the horizon \cite{Linde:2005ht}.

The UV cutoff may be directly implemented at the leading order  by considering the states with finite occupation numbers relative to the Bunch-Davies vacuum. Alternatively,
one can impose the cutoff by hand, as for example in \cite{Greene:2004np}, where the high energy states are
cut off by regulating the momenta according to $\partial^2 \to e^{\partial^2/{\cal M}^2 a_0^2} \partial^2$, yielding a factor $e^{-k^2/{\cal M}^2 a_0^2}$ in the integrand in \eqref{delta T}. In that case, one can consider larger expectation values of the `bare' number operators, with the physical expectation values nevertheless suppressed by the explicit 
nonlocal cutoff.

The terms in \eqref{greenfunction} involving $e^{-ik(2\eta_0 - \eta_1 - \eta_2)}$ will approximately integrate to zero, and so the leading backreaction effect is actually given by the second order corrections to the state:\footnote{The first order change considered by Porrati \cite{Porrati:2004gz, Porrati:2004dm} is localized to the boundary, and so, as explained in \cite{Greene:2004np}, there are renormalization ambiguities, in addition to it being negligible in the bulk.} 
\begin{align}
\delta \langle b | T^0_0 | b \rangle &= -\frac{3\pi^2 a_0^6 H^6}{2(2\pi)^3 a^2} \int \mathrm{d}k\, |\delta\kappa|^2 k^{-5} (1+k^2\eta_0^2)^2 (5 + 2 k^2 \eta^2) e^{-k^2/{\cal M}^2 a_0^2} \nonumber \\
&\approx - \frac{3 a_0^2}{8\pi a^4} \int \mathrm{d}k\, |\delta\kappa|^2 k e^{-k^2/{\cal M}^2 a_0^2} \, , \label{backreaction}
\end{align}
where the last line is of course valid only if $\delta\kappa$ is sufficiently blue so that the integral is dominated by the UV contributions. The explicit cutoff has been retained for completeness and illustrative purposes. 
One sees that the backreaction is UV finite if $\delta\kappa \sim k^{-(1+\epsilon)}$ or more suppressed for large $k$. In this case, we can drop the exponential cutoff in (\ref{backreaction}), essentially by taking the limit ${\cal M} \rightarrow \infty$. If on the other hand the 
function $\delta \kappa$ has less intrinsic suppression for large momenta $k$ it is essential to keep the cutoff 
${\cal M}$ finite. This scaling behavior, as we will see below,
is crucial to ensure the perturbativity of the theory in the covariant local limit. We will return to this in what follows.  
In any case, one sees that the backreaction redshifts away like $a^{\gamma}, 2 \le \gamma \le 4$.

This represents a finite renormalization of the stress-energy tensor which corrects the background via its gravitational effects. Specifically,
in this state the effective Friedman equation with the backreaction included is $H^2 = \frac{1}{3 M_\text{Pl}^2}\left( \rho^{(0)} + \delta \langle b | T^0_0 | b \rangle \right)$. Using it we can derive three constraints on the scale of backreaction:
\begin{itemize}
\item Demanding perturbativity yields 
\begin{equation}
\Delta \equiv \delta \langle b | T^0_0 | b \rangle \ll \rho^{(0)} \approx 3 H^2 M_\text{Pl}^2 \, .
\end{equation}
\item Requiring validity of the slow roll leads to
\begin{equation}
\epsilon = - \frac{\dot{H}}{H^2} = -\frac{\dot{\rho}}{2H \rho} \approx \frac{\dot{\Delta}}{6 H^3 M_\text{Pl}^2} + \epsilon_\text{measured} \sim \epsilon_\text{measured} \implies \Delta \lesssim \frac{3}{2} H^2 M_\text{Pl}^2 \epsilon_\text{measured} \, . \label{slow roll}
\end{equation}
\item Finally, retaining the scale invariance of the scalar power spectrum implies
\begin{align}
&~~ \frac{\ddot{\rho}}{H^2 \rho} \approx \frac{\ddot{\Delta}}{3 H^4 M_\text{Pl}^2} + \epsilon_\text{measured}(\epsilon_\text{measured} + \eta_\text{measured}) \nonumber \\
&\implies \Delta \lesssim \frac{3}{16} H^2 M_\text{Pl}^2 \epsilon_\text{measured}(\epsilon_\text{measured} + \eta_\text{measured}) \, . 
\label{second slow roll}
\end{align}
\end{itemize}
The third constraint yields the strongest bounds on the initial state. However, this constraint really only applies when modes we can observe on the CMB leave the horizon, and becomes equivalent to the second one within a few efolds. Again, the boundary action formalism leads quickly to these results.

\subsection{Correcting Observables}

Now we can check the scaling of the corrected power spectrum in an excited state. The power spectrum of fluctuations is
\begin{align}
P_b(k) &= \lim_{\eta \to 0^-} \frac{k^3}{2\pi^2} G_b(k; \eta,\eta)  = P_\text{BD}\left( 1 - 2a_0^3 \left( \delta\kappa \varphi_+(\eta_0)^2 \right) + \mathcal{O}(\delta\kappa^2) \right) \nonumber \\
&= P_\text{BD}\left( 1 + \frac{a_0^3 H^2}{k^3} \mathrm{Im}\left( \delta\kappa (1 - k^2\eta_0^2 + 2ik\eta_0) e^{-2ik\eta_0} \right) + \mathcal{O}(\delta\kappa^2) \right) \, .
\end{align}
In \cite{Schalm:2004qk, Greene:2004np, Porrati:2004gz, Porrati:2004dm, Nitti:2005ym, Jiang:2015hfa, Jiang:2016nok} the change of a state was due to an irrelevant boundary operator, which in turn was due to integrating out unknown high energy physics. This suggests the following parametrization, involving an explicit momentum dependence which suppresses the deformations in the UV in the construction of the initial state:
\begin{equation}
\delta\kappa = \beta_n M \left( \frac{k}{a_0 M} \right)^n\, .  \label{delta kappa}
\end{equation}
Here $M$ is the UV regulator, and $\beta_n$ is a dimensionless coefficient. In principle $M$ could be different from the cutoff ${\cal M}$ of the previous section. Note that demanding $n<-1$ ensures the finiteness of the stress-energy tensor in the excited state even without the nonlocal cutoff, as we have seen in Eq. (\ref{backreaction}). 
We will mostly work with this requirement, although we will reflect on the situation where the suppression of $\delta\kappa$ is weaker and the explicit nonlocal cutoff $\exp(\partial^2/a_0^2 {\cal M}^2)$ is used. The correction to the power spectrum is
\begin{align}
\frac{\delta P}{P_\text{BD}} = \left(\frac{H}{M}\right)^{n-1} \Bigg[ &\left(\frac{k}{a_0 H}\right)^{n-3} \left( \mathrm{Re} \beta_n \sin\left( \frac{2k}{a_0 H} \right) + \mathrm{Im} \beta_n \cos\left( \frac{2k}{a_0 H} \right) \right) \nonumber \\
+ 2&\left(\frac{k}{a_0 H}\right)^{n-2} \left( \mathrm{Re} \beta_n \cos\left( \frac{2k}{a_0 H} \right) + \mathrm{Im} \beta_n \sin\left( \frac{2k}{a_0 H} \right) \right) \nonumber \\
- &\left(\frac{k}{a_0 H}\right)^{n-1} \left( \mathrm{Re} \beta_n \sin\left( \frac{2k}{a_0 H} \right) + \mathrm{Im} \beta_n \cos\left( \frac{2k}{a_0 H} \right) \right) \Bigg] \, . \label{power spectrum change}
\end{align}

The powers of $a_0$ above show that the 
correlation functions at a later time receive only a small correction compared to their values in the Bunch-Davies state. They are redshifted as momenta due to de Sitter expansion and covariance. Since the power spectrum essentially measures the modes as they leave the horizon, the correction to modes that were deeply subhorizon at $\eta_0$ are greatly suppressed. From \eqref{power spectrum change} one has
\begin{equation}
k \gg a_0 H \implies \frac{\delta P}{P_\text{BD}}\bigg|_{k = a H} \sim \beta_n \left(\frac{a}{a_0} \frac{H}{M} \right)^{n-1} \, , 
\label{large k power spectrum change}
\end{equation}
which is already exponentially (in time) suppressed (by powers of $a/a_0 \simeq \exp(Ht)$) for $n < 1$. As noted above, the condition
$n<-1$ guarantees the finiteness of the corrected stress-energy tensor. 

This conforms with the fact that the standard approach
to QFT fails in states which are separated from the Bunch-Davies vacuum by infinite occupation numbers, such as $\alpha$-vacua \cite{Allen:1985ux, Mottola:1984ar, Danielsson:2002kx, Danielsson:2002qh}. Some of the problems are discussed in \cite{Kaloper:2002cs,Kaloper:2002uj,Starobinsky:2002rp, Shukla:2016bnu}. If the initial state is an excitation of the Bunch-Davies vacuum described by a low energy theory with heavy states and higher dimension, irrelevant operators integrated out, the UV completeness of the theory implies that in a general excited state
the occupation numbers are finite, and the perturbation theory is meaningful \cite{Burgess:2002ub,Kaloper:2003nv}. 
These are the states with $n < -1$ on which we have been mainly focusing. 
Alternatively using the nonlocal cutoff $\exp(\partial^2/a_0^2 {\cal M}^2)$ \cite{Greene:2004np}, one may be able to sharply suppress the 
UV effects regardless of their origin. This 
shows that the seeming enhancements for $n \geq -1$ are merely an illusion, since they are unphysical. If the cutoff is explicitly inserted, the exponential suppression due to it completely overwhelms any power law enhancement with $k$ for $n \geq -1$, for ${\cal M} < M_\text{Pl}^2$, to the point
where such excitations are completely irrelevant in the UV. However, our analysis will establish that this arises naturally by the evolution of the initial state towards the Bunch-Davies vacuum, and is not an arbitrary assumption.

Let us confirm this by computing the occupation numbers in an excited state. 
The number of particles per unit volume, compared to the Bunch-Davies vacuum, is
\begin{align}
N &= \int \frac{\mathrm{d}^3 k}{(2\pi)^3} \langle b | A^\dagger (k) A(k) | b \rangle =  \int \frac{\mathrm{d}^3 k}{(2\pi)^3} \frac{|b|^2}{(1-|b|^2)^2}
\nonumber  \\
&= \int \frac{\mathrm{d}^3 k}{(2\pi)^3} a_0^6 |\delta\kappa|^2 |\varphi_+|^4 \left| \frac{1 - i a_0^3 |\varphi_+|^2 \delta\kappa}{1 + 2 a_0^3 |\varphi_+|^2 \mathrm{Im}(\delta\kappa)} \right|^2.
\end{align}
The large $k$ behaviour of the integrand is
\begin{equation}
N_k \sim 
\begin{cases}
\left|\frac{\delta\kappa}{k}\right|^2 \left( 1 + \left|\frac{\delta\kappa}{k}\right|^2 \right) \, , & \text{ if } |\mathrm{Im}(\delta\kappa)| \ll |\mathrm{Re}(\delta\kappa)| \, ,\\
\left|\frac{\delta\kappa}{k}\right|^2 \, , & \text{ otherwise} \, .
\end{cases}
\end{equation}
In either case we see that $\delta\kappa \sim k^n$ with $n < -\frac{1}{2}$ ensures that the number of particles in the UV is finite. Thus, physically
realistic cases do indeed obey the no-hair scaling, and a nonlocal cutoff is not required. Note that this condition is weaker than the requirement that the extra energy density is finite, $n < -1$, following from \eqref{backreaction}. However it is still stronger than the requirement that the change to the power spectrum decays for large $k$ 
($n < 1$). The main reason for this is that the backreaction and particle number involve integrals of $\delta\kappa$ which smear out the
momentum dependence, whereas the local operator for the power spectrum does not. The power spectrum is thus the least sensitive observable to the UV distortions of the
theory, and one can easily err by picking for initial states such configurations where the power spectrum might not be affected too much, even if the state itself is unphysical.

\subsection{Excitations as Squeezed Bunch-Davies}

So far we have been referring to the non-Bunch-Davies initial states as `Bunch-Davies excitations' in a somewhat heuristic way, basically taking it on faith that the identification is correct. The observables computed in these states indeed support this. Further, this provides a direct link between the boundary action parametrization of initial excitations and a direct second-quantized framework, yielding immediately the systematic normalization of the latter.

We do this by actually constructing these states as normalizable deformations of Bunch-Davies. To this end, we deploy the general 
formalism found in \cite{DeWitt:1975ys}, with the result
\begin{equation}
\langle k_1, \hdots, k_n | b \rangle =
\begin{cases}
i^{n/2} \langle 0 | b \rangle \sum_P f(k_1,k_2) \cdots f(k_{n-1},k_n) \, ,  \qquad & n \text{ even} \, , \\
0 \, , & n \text{ odd}\, .
\end{cases}
\end{equation}
The summation is over the $\frac{n!}{2^{n/2} (n/2)!}$ ways of forming pairs from $\{k_1, k_2, \hdots k_n\}$. In terms of the Bogoliubov coefficients which relate the operators which annihilate $|b\rangle$ to those which annihilate $|0\rangle$ one has
\begin{equation}
f(q,p) = -i \int \mathrm{d}^3k\, {\alpha^*}^{-1}(p,k) \beta^*(k,q) \, .
\end{equation}
In this case the Bogoliubov transformation is `diagonal', with $\alpha(k,k') = \alpha_k \delta^{(3)}(k - k')$ and $\beta(k,k') = \beta_k \delta^{(3)}(k + k')$. Reading off the coefficients from \eqref{modified state} we find
\begin{align}
|b\rangle &= \langle 0 | b \rangle 
\sum_{n=0}^\infty \frac{1}{2^n n!} \int \mathrm{d}^3k_1 \cdots \mathrm{d}^3 k_{n} \, b(k_1)^* \cdots b(k_n)^* | k_1, -k_1, k_2, -k_2, \hdots, k_{2n}, -k_{2n} \rangle \, , \nonumber \\
&= \langle 0 | b \rangle 
\exp \left( \frac{1}{2} \int \mathrm{d}^3k \, b(k)^* a^\dagger(k) a^\dagger(-k) \right) | 0 \rangle \, . 
\label{trafo}
\end{align}
where the second line identifies the operator which turns the Bunch-Davies vacuum into the modified state. Note that this is 
precisely a squeezed state on top of the Bunch-Davies vacuum, akin to the $\alpha$-vacua \cite{Allen:1985ux, Mottola:1984ar, Danielsson:2002kx, Danielsson:2002qh}, but with an explicit UV suppression
manifest in the function $b(k)$. This ensures that the occupation number is finite, and that the transformation (\ref{trafo}) is
normalizable. 
Indeed, using Eqs. \eqref{delta kappa} and \eqref{b coefficient} we see that for $n < 1$
\begin{equation}
b \approx \frac{-\frac{i}{2} \beta_n \left( \frac{k}{a_0 M} \right)^{n-1} \mathrm{e}^{-2i \frac{k}{a_0 H}}}{1-\frac{i}{2} \beta_n \left( \frac{k}{a_0 M} \right)^{n-1}} \approx 
\begin{cases}
\mathrm{e}^{-2i \frac{k}{a_0 H}} \, , &k \ll a_0 M \, , \\
-\frac{i}{2} \beta_n \left( \frac{a_0 M}{k} \right)^{1 - n} 
\, , \qquad & k \gg a_0 M \, .
\end{cases}
\label{cutoffb}
\end{equation}
Not surprisingly by now, we see yet again that the integrals in (\ref{trafo}) will be finite for $n < -1$, as evidenced by the suppression of $b$ in the last line of (\ref{cutoffb}). Without such behavior we would have to explicitly cut off the momentum integral in (\ref{trafo}) to keep the transformation normalizable.

\section{Quantum Inflation}

Inflation in its standard form is a mechanism to get rid of initial inhomogeneities and anisotropies. Once the cosmological constant-like
source starts to dominate the expansion, (almost) everything else dilutes away \cite{Wald:1983ky, Barrow:1985, Barrow:1986, Jensen:1986vy, Moss:1986ud, Kleban:2016sqm}.
Intuitively---by invoking Ehrenfest's theorem---this must also happen with initial quantum excitations of the Bunch-Davies vacuum.
Beside proving this, we will also compute precisely how many efolds it takes to suppress the effect of the 
initial excitations on the observable predictions of inflation to below their intrinsic dynamical value, obtained by the computation in the Bunch-Davies
state. In other words, we can precisely state the `cost' of the choice of Bunch-Davies as the initial state of inflation in terms of the number
of efolds it takes prior to the last 60 efolds in order to iron out generic initial excitations.

So let us take as the initial state a deformation given by \eqref{delta kappa}, with $n < -1$ so that the quantum correction of the stress-energy tensor is finite. In this case we can drop the exponential cutoff factor from \eqref{backreaction}. Then, direct computation yields
\begin{equation}
\Delta(a) = \frac{3}{16 \pi} \left[ |\beta| \left(\frac{H}{M} \right)^{n-1} \right]^2 H^4 \left[ \frac{5}{2}c(n-1) \left(\frac{a_0}{a} \right)^2 + c(n) \left(\frac{a_0}{a} \right)^4 \right] \, ,
\end{equation}
where $c(n) = \frac{1}{n-1} + \frac{2}{n} + \frac{1}{n+1}$. Let the parameters $\beta$ and $M$ be such that at $a = a_0$ the extra stress-energy contribution is just large enough to disrupt slow-roll. Clearly this is the maximal value allowed at the beginning of the slow roll phase, by the fact that the
slow roll is attractor for a sufficiently flat inflationary potential. Eq. \eqref{slow roll} then yields
\begin{equation}
|\beta| \left(\frac{H}{M} \right)^{n-1} = \frac{M_\text{Pl}}{H} \sqrt{\frac{8 \pi \epsilon}{\tilde{c}}} \, , \label{backreaction substitution}
\end{equation}
where $\tilde{c} = |\frac{5}{2} c(n) + c(n-1)|$. Using this we can calculate the change to the power spectrum at a later time. Using \eqref{large k power spectrum change}, we find
\begin{equation}
\frac{\delta P}{P_\text{BD}}\bigg|_{k = a H} \simeq \left( \frac{a_0}{a} \right)^{1-n} \frac{M_\text{Pl}}{H} \sqrt{\frac{8 \pi \epsilon}{\tilde{c}}} < \left( \frac{a_0}{a} \right)^{1-n} \frac{M_\text{Pl}}{H} \sqrt{8 \pi \epsilon} \, .
\label{deformation}
\end{equation}
The inequality holds for $n \gtrsim -13$. For larger negative values of $n$ the dimensionless factor can be larger,
but the redshift suppression $\sim (a_0/a)^{1-n}$ is so efficient that these cases are essentially ignorable.

Indeed, while the contribution of the initial excitations to the scalar power spectrum (\ref{deformation}) may be very large initially, since $\frac{M_\text{Pl}}{H} \gtrsim 10^5$, the correction  dilutes at least as quickly as $1/a^{2}$ because $n<-1$. The number of efolds to reduce the initial ratio 
to a desired value is
\begin{equation}
N = \frac{1}{1-n} \ln \left( \frac{M_{\rm Pl}}{H} \sqrt{8\pi \epsilon} \left(\frac{\delta P}{P} \right)^{-1} \right) \, . \label{no hair efolds 1}
\end{equation}
Thus to suppress the contribution from the initial condition to $\delta P$ to be below $\sim 10^{-2}$ of the Bunch-Davies value $P_{BD}$, using 
the slow roll parameter $\epsilon \sim 0.01$ which allows ${\cal O}(100)$ efolds of inflation, one has
\begin{equation}
N \simeq \frac{1}{1-n} \left( 4 + 2 \log_{10} \left( \frac{M_\text{Pl}}{H} \right) \right) \, . \label{no hair efolds}
\end{equation}
For the highest scale inflation, $H/M_\text{Pl} \sim 10^{-5}$, and the most UV sensitive distribution of excitations,
$n \sim -1$, this yields 
\be
N \simeq 7 \, .
\ee
For lower scale inflation, $N$ should be larger, but generically it will not take much more than $N \sim {\cal O}(10)$ efolds to reduce the initial excitations to the level where they affect scalar perturbations by less than a percent, at which point they are completely negligible.

Remarkably, this completely confirms the intuition that the erasure of the initial distortions of the quantum vacuum during inflation takes about
10 or so efolds \cite{Kaloper:2002cs, Kaloper:2002uj}. We should still verify that the third of the backreaction constraints is obeyed.
However this is straightforward. From \eqref{second slow roll} one sees that this will be satisfied when $\Delta(a)$ has decreased by $\sim \frac{\epsilon}{8} \sim 10^{-3}$ which happens after $\sim 3$ efolds, i.e. even sooner.

Note that the cost of preparation of the initial state as measured by the number of efolds that it takes to erase the excitations 
increases as the scale of inflation is lowered. The reason is that while the background energy density is $\sim H^2 M_\text{Pl}^2$, the 
perturbation to the energy density is $\Delta \sim H^4$, because it arises from relativistic inflaton fluctuations at subhorizon scales.
Therefore, the requirement that the perturbation is just about to disrupt slow roll when $a = a_0$ means that the size of the perturbation as measured by $\beta$ or $M$ must be larger for smaller $H$. In other words, a relatively larger initial excitation is allowed. This is
why it takes longer for this perturbation to wash away.

\subsection{Non-Gaussianities}

The no-hair theorems in de Sitter space \cite{Marolf:2010zp, Marolf:2010nz, Hollands:2010pr} proscribe all hair---not just that endowing the two point function. This also includes
non-Gaussian signatures, which is indeed why they are small during inflation \cite{Maldacena:2002vr}. Nevertheless,
it is interesting from a phenomenological point of view to consider whether the effects induced by initial excitations on higher point correlators 
might be more resilient to inflationary washout. In other words, does one need more efolds than \eqref{no hair efolds} to bleach nonlinear hairs?

To test this, we turn to the three-point function of the perturbation field. The effect on the three-point function from non-Bunch-Davies initial states has been calculated in \cite{Holman:2007na, Flauger:2013hra}. We can translate the result of \cite{Holman:2007na} into our language, and interpret it in the context of the no-hair theorem.  The relative change in the three-point function of the scalar curvature perturbation $\zeta$, in the high frequency limit where gravity is decoupled, yields
\be
\delta_3 \equiv \frac{\delta \langle \zeta_{k_1} \zeta_{k_2} \zeta_{k_3} \rangle}{\langle \zeta_{k_1} \zeta_{k_2} \zeta_{k_3} \rangle} \simeq 
\frac{k_t}{4 \sum_i k_i^{-2}} a_0^3 H^2 \sum_i \frac{\left( 1+ i k_i \eta_0 \right)^2 \mathrm{e}^{-2i k_i \eta_0} \left( 1 - \mathrm{e}^{i \tilde{k}_i \eta_0} \right) \delta\kappa^* + \mathrm{c.c.}}{k_i^5 \tilde{k}_i} \, , \label{delta3}
\ee
where $k_t = \sum_i k_i$, and $\tilde{k}_i = k_t - 2 k_i$. Due to $\tilde{k}_i$ in the denominator, $\delta_3$ is maximized in the so-called `folded limit' defined by taking $\tilde{k}_i \to 0$ for some $i$. This case is interesting as other effects which alter the three-point function do not display this signature (occurring instead in the 
equilateral and squeezed limits). Thus, any unusual behavior in the folded limit might indicate the significance of a non-Bunch-Davies initial state.

Let us now write $\delta \kappa$ as in \eqref{delta kappa}, and use the backreaction constraint to eliminate the $\beta$ and $M$ parameters via \eqref{backreaction substitution}. As before we take generic phases of $\beta$, 
assuming $\mathrm{Re}(\beta) \sim \mathrm{Im}(\beta) \sim |\beta|$. In the folded limit of \eqref{delta3},
setting $k_2 = k_3 = \frac{1}{2}k_1 = \frac{1}{2}k$, the dominant contribution is
\be
\delta_3 \simeq \frac{1}{18} \left( \frac{M_\text{Pl}}{H} \right) \sqrt{\frac{8 \pi \epsilon}{\tilde{c}}} \left( \frac{k}{a_0 H} \right)^n \simeq \frac{1}{18} \left( \frac{k}{a_0 H} \right) \frac{\delta P}{P_\text{BD}} \, .
\ee
For a mode leaving the horizon at $k = a H$, this is larger than the fractional change in the power spectrum by a factor of $a/a_0$. However, the constraints on the primordial contribution to the three-point function are much less severe than those on the two-point function. So a value of $\delta_3$ which is larger than $\frac{\delta P}{P_\text{BD}}$ is still practically irrelevant. In particular, using \eqref{no hair efolds 1}, we can write
\begin{equation}
\delta_3 \simeq \frac{1}{18} \left( \sqrt{\frac{\tilde{c}}{8\pi\epsilon}}\frac{H}{M_\text{Pl}} \frac{\delta P}{P_\text{BD}} \right)^{-\frac{1}{1-n}} \frac{\delta P}{P_\text{BD}} \, ,
\end{equation}
and so for $H \sim 10^{-5} M_\text{Pl}$, when the fractional change in the power spectrum is $1\%$ the fractional change in the three-point function for $n = -1$ (which gives the largest effect) is at most $\mathcal{O}(1)$. Thus the 
non-Gaussianities induced by initial conditions will be suppressed down to the level of intrinsic dynamical non-Gaussianities---induced by scattering in the Bunch-Davies vacuum---by the same $\mathcal{O}(10)$ efolds which prepared the Bunch-Davies state. Any longer stage of inflation prior to the last 60 efolds will suppress them even more. This is in agreement with the analysis done in \cite{Flauger:2013hra}.

Furthermore, as noted in \cite{Holman:2007na}, the extra factor of $\frac{k}{a_0 H}$ in $\delta_3$ compared with the modification of the scalar power spectrum essentially gets washed out once one considers the projection of the three-point function onto the two-dimensional surface of last scattering. In this case the change in the bispectrum would be irrelevant at the same time as the change in the power spectrum.\footnote{This `information suppression' might be avoided by fully three-dimensional observation, such as large scale structure surveys.}

\section{Vacua and Bouncing Universes}

We can raise similar questions about the evolution of quantum excitations in the so-called bouncing universes. These models rely on a period of contraction prior to a bounce and eventual re-expansion as a possible alternative to the generation of the inflationary perturbations (see, e.g., \cite{Battefeld:2014uga} for a review). 
It has been pointed out that such scenarios suffer from severe fine-tuning problems related to their classical initial conditions (for example see \cite{Turner:1997ih, Kaloper:1998eg, Tolley:2007nq, Levy:2016xcl}). These prompted the proposal for cyclic universes where the classical fine-tuning problems are addressed by repeated incarnations of the universe (see \cite{Steinhardt:2001vw, Linde:2002ws} for some details of the idea and its critique). Some aspects of the quantum mechanics of fluctuations in the bouncing analogue of the Bunch-Davies state have been analyzed (see e.g. \cite{Gratton:2003pe, Boyle:2004gv}); however, a general discussion of how the Bunch-Davies state is attained appears to have been overlooked to date. Without getting into a detailed discussion of the bounce dynamics (and what may or may not cause it), we wish to merely point out the difference between quantum state selection 
and evolution in inflation and in bouncing universes. While in inflation the Bunch-Davies state is an attractor, it is far less clear whether this is true in
bouncing cosmologies. Hence, such models need to be explored in more detail to see if a quantum attractor mechanism exists, that would justify the calculations of the perturbations in the Bunch-Davies state.

To make things more precise, we consider a flat FRW universe with scale factor $a = a_0 (t/t_0)^p = a_0 (\eta/\eta_0)^{\frac{p}{1-p}}$, with $\eta \in (-\infty, 0)$ and $0 < p < 1$. The mode functions for a massless scalar in the Bunch-Davies vacuum are
\begin{equation}
\varphi = \frac{\sqrt{-\eta}}{a(\eta)} \sqrt{\frac{\pi}{4}} \mathrm{e}^{i \frac{\pi}{4} (2\nu +1)} H^{(1)}_\nu (-k \eta) \, ,
\end{equation}
where $H^{(1)}_\nu$ is the Hankel function of the first kind, and its order is $\nu = \frac{1-3p}{2(1-p)}$. Note that the case of an expanding de Sitter universe can be formally reached by taking the limit $p \to \infty$, or $\nu \rightarrow 3/2$. On the other hand, the phase which is argued to be crucial for scale invariance in bouncing cosmologies, corresponding to a very stiff equation of state $P/\rho \gg 1$ \cite{Gratton:2003pe, Boyle:2004gv}, yields $p \rightarrow 0$, and so $\nu \rightarrow 1/2$.

Let us now consider what happens if the initial state is not exactly the Bunch-Davies one. 
The boundary action formalism can be readily applied to this situation. 
Note that this state describes the universe well before the
bounce stage, and in principle it should be describable by means of a standard EFT. Thus, 
by Eq. \eqref{backreaction} the change in the energy density due to a deviation from the vacuum is
\ba
&&\Delta(\eta) = \frac{3\pi}{128} \left( \frac{1-p}{p} \right)^2 H_0^2 \left(\frac{a_0}{a} \right)^2 \int \mathrm{d}(-k\eta_0) |\delta\kappa|^2 (-k\eta_0)^\frac{3-p}{1-p} (-k\eta)^\frac{1-3p}{1-p}  \\
&& ~~~~~~~~~~~~~~~~~~~~~~~~~~~~~~~~~~~~~~~~~ 
\times | H^{(1)}_\nu(-k\eta_0) |^4 \left( | H^{(1)}_\nu(-k\eta) |^2 + | H^{(1)}_{\nu-1}(-k\eta) |^2 \right) + \cdots \, , ~~ \nonumber
\label{deltahank}
\ea
where the ellipsis refers to more terms of a roughly similar form. Note that the integral is UV finite for $\delta \kappa \sim k^n$ with $n < -1$, just as in the inflationary case, thanks to the behavior of Hankel functions for large argument. Indeed, for large argument, $H_\nu(x) \sim e^{ix}/\sqrt{x}$ which means that each power of the Hankel function in (\ref{deltahank}) contributes a power of $1/\sqrt{k}$ to the integrand. Thus at large momenta, the integrand behaves as $\sim k |\delta \kappa|^2$.  Also note that $\Delta$ involves a prefactor $\sim a^{-2}$ and so, unlike in the case of inflation, as time goes on, and the universe shrinks, $\Delta$ increases. The explicit $\eta$ dependence in the integral cannot counter this, as the Hankel functions blow up for small argument, rendering the mode functions finite.
Nevertheless, the background energy density driving the contraction behaves as $\rho \sim 1/a^{2/p}$. Since $p<1$ this grows more quickly than the perturbation $\Delta$, subduing the backreaction. This keeps control of the collapse at the classical level \cite{Gratton:2003pe, Boyle:2004gv}. 

The question is: what happens with the corrections to the observables due to the excitations ensconced in the initial quantum state 
of collapse, when it is not the Bunch-Davies vacuum? Writing $\delta\kappa$ as \eqref{delta kappa}, at the initial time $\eta_0$ one has
\begin{equation}
\Delta(\eta_0) = \frac{3\pi}{128} I(p,n) \left( \frac{1-p}{p} \right)^2 H_0^2 \beta^2 M^2 \left( \frac{H_0}{M} \frac{1-p}{p} \right)^{2n}\, , \label{contracting backreaction}
\end{equation}
where $I(p,n) = \int_1^\infty \mathrm{d}x \, x^{4+2n} |H^{(1)}_\nu(x)|^4 \left( | H^{(1)}_\nu(x) |^2 + | H^{(1)}_{\nu-1}(x) |^2 \right) + \cdots$\, .
Using \eqref{contracting backreaction} to constrain $\beta$ and $M$ in terms of the initial backreaction, we can now examine the change in the power spectrum. We find
\begin{equation}
\frac{\delta P}{P_\text{BD}} \bigg|_{-k \eta = 1} \simeq \sqrt{\frac{32}{3 \pi I(p,n)}} \frac{\Delta_0^{\frac{1}{2}}}{H_0^2} \left( \frac{a}{a_0} \right)^{-n\frac{1-p}{p}} \left| H^{(1)}_\nu \left( \left(\frac{a_0}{a}\right)^\frac{1-p}{p} \right) \right|^2 \, . \label{contracting power spectrum change}
\end{equation}
So the fractional change in the power spectrum decreases for modes which leave the horizon at a later time, just as in the inflationary case considered above (recall that we are considering $n < -1$ and $0 < p < 1$).

So far so good. However, let us pay closer attention to (\ref{contracting power spectrum change}) in the limit relevant for bouncing universe models $p \rightarrow 0$, which is argued to be required in order to retain the scale
invariance of perturbations \cite{Gratton:2003pe, Boyle:2004gv}. In this case, 
$a \rightarrow {\rm const}$ and $H = \dot a/a \sim p \rightarrow 0$, such that $H_0/p$ is finite. Thus the backreaction remains under control. Let us therefore fix it by requiring $\Delta_0$ to be a small fraction of the initial background energy density, $\sim M_\text{Pl}^2 H_0^2$. This implies that \eqref{contracting power spectrum change} involves an overall factor $M_\text{Pl}/H_0$, just as in the inflationary case. But because $H_0 \rightarrow 0$ in this phase, $M_\text{Pl}/H_0 \gg 1$, much more so than in the case of inflation. 
Because of this, bouncing universes can accommodate much larger initial excitations that can distort the scalar power spectrum much more. In principle, they could even be as large as unity, or even more, implying that the whole approximation of the background universe by a homogeneous FRW metric is invalid. Such initial quantum distortions need to be suppressed. In inflation this occurs automatically once accelerated expansion starts. While
the background backreaction in bouncing universes remains under control, the influence of initial excitations on observables is more persistent. In the absence of a more efficient mechanism to smooth these excitations away bouncing universes seem to require fine-tunings by hand to pick the right form of the perturbations from generic initial conditions. 
In a way, the Bunch-Davies vacuum state seems to be a much weaker attractor. 

Note that we have used initial backreaction as a measure of the excitations away from the Bunch-Davies state. One could try imposing 
direct limits on $\beta_n$ in \eqref{delta kappa} instead. In such an approach one opens the door to sensitivity to UV physics,
since generically the results will depend explicitly on the UV cutoff $M$ because the backreaction and the modification of the power spectrum 
involve positive powers of the ratio $M/H_0$. This would therefore seem counterproductive, requiring direct tunings of UV physics  in ways that conflict with 
decoupling. Fixing the cutoff $M$ leads to the same problems, as does fixing the ratio $M/H_0$. 
The bottom line is that in the absence of the inflationary redshifts one needs to develop different mechanisms
that may even have to go beyond standard field theory to justify using the Bunch-Davies state to compute perturbations in bouncing cosmologies.

\section{Summary}

The standard calculation of inflationary perturbations involves the computation of the spectrum of fluctuations of
relativistic fields in the Bunch-Davies vacuum. On the other hand, one can---and should---imagine that the initial quantum state of the universe in the beginning of inflation is more general, and check the effects of the more general choices on the observables. After all inflation is the mechanism for smoothing the initial conditions away using accelerated expansion as the attractor dynamics.

At the quantum level, the same phenomenon reoccurs: 
inflationary expansion induces large redshift factors in the expectation values of observable operators in generic initial states which rapidly diminish the effects of initial excitations. The quantum cosmic no-hair theorem picks
the Bunch-Davies state as the vacuum, and evolution turns it into the attractor. The underlying physics of the quantum balding of an initial state is just decoupling, whereby the IR observables are insensitive to UV effects due to the large relative redshifts. Using the  boundary action formalism to determine the magnitude of initial excitations due to their backreaction on the background, we found that the effects of initial excitations reduce to insignificant levels within ${\cal O}(10)$ efolds. 
This occurs for both the scalar power spectrum and non-Gaussianities. If inflation involves ${\cal O}(10)$ efolds preceding the last 60, the quantum effects of initial conditions are wiped out. Our results confirm that inflationary quasi-de Sitter expansion is indeed the mechanism which smooths out the universe at both the classical and quantum levels. 
The `thermalization rate'  of Eq. \eqref{no hair efolds} (in the sense of the initial excitations being reduced below the level of the quantum quasi-de Sitter fluctuations) is quite rapid. Note that while it has been colloquially said that inflation prepares the Bunch-Davies state as the vacuum of fluctuations, the precise and general details were lacking in the literature. Our work fills that gap. 
We nevertheless should note that larger signatures may occur in ``just-so" models where inflation lasts only ${\cal O}(60)$ efolds, but this may require fine-tuning both the theory and the initial state from the model building point of view. Some recent examples are \cite{Albrecht:2014aga, Kanno:2015lja}.

For bouncing universes, the tuning of the initial quantum state has not previously been discussed in the literature. In this case the dynamics is different; as a result, the EFT description permits much larger initial excitations which are much harder to suppress. The resultant Bunch-Davies state typically used for computing the perturbations is a much weaker attractor. Hence if one starts with generic initial conditions in the collapsing phase, one has to pick more carefully the right initial state as a function of the duration of collapse to get the observed quantities. By itself, this is a fine tuning. A dynamical mechanism explaining it would be preferable. Perhaps this could be alleviated in cyclic universes \cite{Steinhardt:2001vw} where the universe undergoes many stages of collapse followed by expansion, with a late long phase dominated by a small cosmological constant. In this case the `bleaching' of the quantum state of the universe to a Bunch-Davies vacuum would be done by a long late low scale inflation \cite{Felder:2002jk}. This does not seem like a very economical scenario since the 
dilution requires very long times (see e.g. \cite{Kaloper:2004gp}). 
But if the number of cycles is large\ldots{}
A more careful investigation would seem to be warranted to test these issues.

We note in closing that our analysis does not preclude large effects due to the quantum initial conditions on the post-inflationary universe. It does make them very unlikely however, if inflation is longer than the minimal 60 efolds. 
If for example we consider large field slow roll inflation, it can easily last longer than 60 efolds, by at least ${\cal O}(10)$ efolds, and as we saw the initial excitations of the quantum state of the universe will be suppressed to very small levels. Similarly, in the case of false vacuum inflation, where the last stage happens after the tunneling out of a false vacuum (see, e.g., \cite{Linde:1999wv, Freivogel:2005vv}), the quantum state at the onset of the last stage is prepared to be the Bunch-Davies one by the long time the universe lingered in the false vacuum. Again, the quantum excitations will be grossly suppressed.  To avoid this, one needs a `just-so' inflation: the last 60 efolds need to start with a `bang' which provides a large initial sudden excitation, disrupting the adiabaticity of cosmic evolution \cite{Tanaka:2000jw, Kaloper:2003nv}. While perhaps possible, such dynamics seem to be tuned.

\section*{Acknowledgements}
We would like to thank Albion Lawrence for useful discussions.
N.K. thanks the CERN Theory Division for hospitality in the course of this work. This work is supported in part by DOE Grants DE-SC0009999 and DE-SC0019081. 

\bibliographystyle{JHEP}
\bibliography{nohairinflation}

\end{document}